\documentclass{emulateapj}

\slugcomment{To be published in Astrophysics Journal Letters} 

\shorttitle{Chandra/HETGS Observations of Capella}
\shortauthors{Ishibashi et al.}

\begin{document}

\title{{\it Chandra/HETGS} Observations of the Capella System: \\
            the Primary as a Dominating X-ray Source }

%
%
\author{Kazunori Ishibashi, Daniel Dewey, David P. Huenemoerder, and Paola Testa}
\affil{Massachusetts Institute of Technology Kavli Institute for 
Astrophysics and Space Research, 77 Massachusetts Ave. Cambridge, MA 02139}

\email{bish@space.mit.edu}

\begin{abstract}

Using the {\it Chandra}/High Energy Transmission Grating Spectrometer (hereafter HETGS) 
we have detected Doppler motion of Capella's X-ray emission 
lines in the 6 -- 25\AA\ wave-band. The observed motion follows the expected orbital 
motion of Capella's primary. This finding implies that the primary G8~{\sc{III}} star, 
not the secondary G1~{\sc{III}} star in the Hertzsprung gap, has been the dominant
source of hot 10$^{6.8}$ -- 10$^{7}$~K plasma at least in the last six years. In addition, 
the results demonstrate the long-term stability of the HETGS and demonstrate small 
uncertainties of 25 and 33 km~s$^{-1}$ in the velocity determination 
with the HEG and MEG, respectively. 

\end{abstract}


\section{Introduction}
Capella ($\alpha$ Aurigae $=$ HD~34029, $\alpha = 
5^{h}~16^{m}~41.4^{s}$, $\delta =$  45\arcdeg~59\arcmin~53\arcsec\ 
in Epoch 2000.0) is a well-studied binary system consisting mainly 
of two cool giants: G8~{\sc{III}} and G1~{\sc{III}} (often labeled as 
Aa and Ab, respectively).  The two giants have very similar masses of 
2.69 and 2.56~M$_{\odot}$ (Hummel et al. 1994)\footnote{It makes little 
difference in the following work if the mass values by Barlow et al.(1993) 
are quoted, instead.}.  The cooler, more massive G8~{\sc{III}} star 
is a core He-burning (CHeB) clump giant, while the hotter secondary G1~{\sc{III}} 
star is a rapidly rotating giant crossing the Hertzsprung gap for the 
first time (Pilachowski \& Sowell 1992; Barlow, Fekel, \& Scarfe 1993 
and references therein). Generally both gap and clump stars show hot 
($T \sim$10$^{7}$K) coronal emission measure distribution, although
gap stars are often found more deficient in X-rays than clump 
stars (Ayres et al. 1998). 

In the case of the Capella system, 
it was not clear how much each star contributed to the total X-ray 
luminosity. Linsky et al. (1998) found that the contribution of 
the G8 star to the total Fe~{\sc{XXI}} $\lambda$1354 flux (forming 
at T $\sim$ 10$^{7}$K) equaled that of the G1 star in the 1995 
and 1996 observations with the {\it Hubble Space Telescope}/Goddard High-Resolution 
Spectrometer, whereas the G8 star contributed less in other
UV lines forming at much lower temperatures. Johnson et al. (2002), instead,  
found in 1999 with the {\it HST}/Space Telescope Imaging Spectrograph 
that the G8 star exhibited a negligible contribution to the same line, 
indicating that the 10$^{7}$K plasma of the G8 star was variable on 
a long time scale. In contrast to the finding by Johnson et al., 
Young et al. (2001) found in 2000 with the {\it Far Ultraviolet 
Spectroscopic Explorer} that the Fe~{\sc{XVIII}}~$\lambda$974.85 line 
(forming at T $\sim$ 10$^{6.8}$K) originated mainly from the G8 star. 
Since the emission measure distribution of Capella sharply peaked 
at T $\sim$ 10$^{6.8}$K (Brickhouse et al. 2000; Ness et al. 2003; 
Argiroffi et al. 2003; and references therein), it seemed natural 
to expect that the main contribution to Capella's X-ray emission 
originated from the G8 star. However, it was not proven definitively.

To date, no X-ray telescope can resolve these stars spatially 
(separation $\approx$ 0.06\arcsec; c.f., Young et al. 2002
on spatially separating the two giants in the UV with the {\it HST}/Faint
Object Camera). However, a definitive determination
of the origin of Capella's hard X-rays can be achieved by 
following the Doppler motion of bright emission lines in the 
observed {\it Chandra}/HETG spectra in its six years' operation.

\begin{deluxetable*}{cccrrrccc}
\tabletypesize{\scriptsize}
\tablecolumns{9}
\tablewidth{0pc}
\tablecaption{{\it Chandra}/HETGS Observations of Capella}
\tablehead{
\cline{1-9} \\
\colhead{ } & \colhead{ }          & \colhead{Good Time}& \colhead{ }      & \multicolumn{2}{c}{Apparent Velocity} & \colhead{Corrected Radial} & \colhead{Barycenter} &\colhead{Spacecraft} \\
\colhead{ } & \colhead{Date Start} & \colhead{Interval} & \colhead{Phase}  &  \colhead{V$_{obs}$} & \colhead{3$\sigma$ confidence}  & \colhead{Velocity}  & \colhead{Correction}  & \colhead{Velocity}\\
\colhead{ObsID} & \colhead{(MJD)}  & \colhead{(ksec)}   & \colhead{$\phi$} & \multicolumn{2}{c}{(km/s)} & \colhead{(V$_{bary}$ km/s)} & \colhead{ (C$_{bary}$ km/s)} &\colhead{(C$_{sc}$ km/s)}}
\startdata
1099  & 51418.329 &  14.6 & 0.398 & $-$34.9 & ($-$19.3 $\sim$ $-$52.7) &$-$8.65& 26.0 & 0.214 \\ 
1235  & 51418.512 &  14.6 & 0.400 & $-$25.3 & ($-$7.77 $\sim$ $-$41.7) &  0.97 & 26.0 & 0.272 \\
1100  & 51418.694 &  14.6 & 0.402 & $-$10.0 & (+8.94 $\sim$ $-$29.5)   &  16.4 & 26.1 & 0.325 \\
1236  & 51418.877 &  14.6 & 0.404 & $-$13.2 & (+6.31 $\sim$ $-$34.6)   &  13.3 & 26.1 & 0.375 \\
1101  & 51419.060 &  14.6 & 0.405 & $-$16.3 & (+4.25 $\sim$ $-$32.2)   &  10.2 & 26.1 & 0.420 \\
1237  & 51419.243 &  14.6 & 0.407 & $-$15.0 & (+5.01 $\sim$ $-$32.2)   &  11.6 & 26.1 & 0.451 \\
1103  & 51445.257 &  40.5 & 0.657 & $-$6.92 & (+3.04 $\sim$ $-$16.4)   &  20.4 & 26.9 & 0.392 \\
1318  & 51446.560 &  26.7 & 0.670 & $-$10.2 & (+4.37 $\sim$ $-$24.4)   &  16.4 & 26.8 & $-$0.182\\
0057  & 51606.687 &  28.8 & 0.209 &  67.4   & (+80.1 $\sim$ +54.7)     &  40.4 & $-$27.3 & 0.352 \\
1010  & 51951.516 &  29.5 & 0.524 &  26.2   & (+42.7 $\sim$ +16.2)     &  1.57 & $-$24.2 & $-$0.468\\
2583  & 52393.741 &  27.6 & 0.775 &  54.6   & (+68.1 $\sim$ +44.6)     &  35.4 & $-$18.8 & $-$0.485\\
3674  & 52909.818 &  28.7 & 0.736 & $-$4.34 & (+9.86 $\sim$ $-$17.1)   &  21.7 & 26.7 & $-$0.623\\
5040  & 53258.924 &  28.7 & 0.092 &  7.10   & (+22.6 $\sim$ $-$5.04)   &  33.6 & 27.3 & $-$0.785\\
5955  & 53457.553 &  28.7 & 0.002 &  65.7   & (+80.3 $\sim$ +49.1)     &  39.2 & $-$26.5 & 0.000 \\
\enddata
\end{deluxetable*}

\section{Observations and Data Processing}

Capella has been observed 14 times (see Table~1) in the Timed Event (TE) mode 
with the {\it Chandra}/HETGS (Canizares et al. 2005) for purposes of calibration
and data collection for the Emission Line Project\footnote{http://cxc.harvard.edu/elp/ELP.html}. 

Each dataset was obtained through the Chandra X-ray Center data 
archive and was reprocessed with the CIAO tools (version 3.2.1); and 
ancillary effective area and grating line response products were generated 
with the CALDB 3.0.1. The reprocessing included the latest updates on 
the ACIS-S chip geometry and the MEG grating period\footnote{
http://space.mit.edu/CXC/docs/docs.html\#acis\_s\_geom}. 
No pixel randomization in the output of \verb|tg_resolve_events| was added
and the default CIAO extraction settings were applied. 

Since the Capella system is a modestly bright X-ray source 
(photon flux~$f_X \approx$~2~cps at 0$^{th}$ order image in the HETG wave-band), 
its 0$^{th}$ order image is piled up. This leads to poor determination 
of the centroid peak in the 0$^{th}$ order image and hence a poorer 
zero-wavelength solution. To remedy this, the 0$^{th}$ order position 
was re-derived based on the intersection of the grating arms and 
the 0$^{th}$-order ACIS frame transfer streak.

\begin{deluxetable}{crr}
\tablecolumns{3}
\tablewidth{0pc}
\tablecaption{Wavelength ranges selected for Doppler measurement}
\tablehead{
\cline{1-3} \\
\colhead{No.} & \colhead{Range $\lambda$ (\AA)}  & \colhead{Selected Emission lines ($\lambda_{lab}$)}}
\startdata
    & \multicolumn{2}{c}{for HEG and MEG}                          \\
    &              &                                               \\
1   & 6.0  -- 6.50 & Si~{\sc{XIV}} $\lambda\lambda$6.1804, 6.1858  \\
2   & 6.6  -- 6.72 & Si~{\sc{XIII}} $\lambda\lambda$6.6479, 6.6882 \\
3   & 8.3  -- 8.55 & Mg~{\sc{XII}} $\lambda\lambda$8.4192, 8.4246  \\
4   & 9.12 -- 9.35 & Mg~{\sc{XI}} $\lambda\lambda$9.1687,9.2312,9.3143\\
    & \nodata      & Fe~{\sc{XXI}} $\lambda$9.1944                 \\
5   & 12.00 -- 12.20&Ne~{\sc{X}} $\lambda\lambda$12.132, 12.137    \\
    & \nodata      & Fe~{\sc{XVII}} $\lambda$12.124                \\
6   & 14.9 -- 15.1 & Fe~{\sc{XVII}} $\lambda$15.014                \\
    & \nodata      & Fe~{\sc{XIX}} $\lambda$15.079                 \\
    &              &                                               \\
    & \multicolumn{2}{c}{for MEG only}                             \\
    &              &                                               \\
7   & 18.85 -- 19.1& O~{\sc{VIII}} $\lambda\lambda$18.967,18.988   \\
8   & 21.4 -- 22.2 & O~{\sc{VII}} $\lambda\lambda$21.602,21.804,22.098 \\
9   & 24.5 -- 25.1 & N~{\sc{VII}} $\lambda\lambda$24.779,24.785    \\
\enddata
\end{deluxetable}

\section{Line Analysis}

To determine accurate Doppler velocities of the bright emission 
lines in the Capella spectrum, we followed this procedure:
(1) select a region of Capella's spectrum containing strong 
emission lines; (2) fit the line position of the $\pm$1$^{st}$ orders 
of the HEG and MEG spectra separately; and then (3) calculate this region's 
Doppler velocity.  The average value of these velocities 
from spectral segments and grating 
orders was taken to be the apparent Doppler velocity of Capella 
for the observation. 

Despite the high spectral resolution obtained with the HETGS, 
almost all of the observed emission lines are blended 
(e.g., all hydrogenic lines are a blend originating from two transitions 
to the ground state, 1s~$^{2}$S$_{1/2}$, from 2p~$^{2}$P$_{3/2}$ and 
2p~$^{2}$P$_{1/2}$ states). Such an unresolved line complex was modeled 
as a blend in order to make a precise measurement of 
its apparent Doppler shift. For the baseline spectral model, 
a three-temperature {\it APED} plasma model\footnote{{\it APED} is a model 
for collisionally excited thermal plasma (Smith et al. 2001).} was chosen 
to represent the Capella spectrum 
(similar to the model used in Canizares et al. 2000). 
The three temperatures were fixed (at $kT\sim$10$^{6.3}$, 10$^{6.8}$, 
and 10$^{7.1}$). The three free parameters, the normalization of line emission 
and the apparent Doppler velocity and width, were then fit by minimizing 
the C-statistic using the Levenberg-Marquardt method as implemented 
in ISIS version 1.3.0  (Houck \& Denicola 2000). All quoted errors 
in the Doppler velocity term  correspond to 3$\sigma$ confidence 
levels (see the fifth and sixth columns in Table~1). 

The regions were selected to include 
bright emission lines; the selected spectral regions and emission
lines included in the fitting process are tabulated in Table~2. 
Once all the measurements were 
done, we took all of the velocities per wavelength region per 
grating, rejected the highest and lowest (i.e., {\it possibly anomalous}) 
data points, and then took the mean as the apparent velocity of the system. 
The measured Doppler widths were small and statistically insignificant
($\leq$150~km~s$^{-1}$).

Each measurement of apparent Doppler velocity needed to be corrected 
for the barycentric motion of the Earth around the center of mass of 
the Solar system. The barycentric correction was 
made using the algorithm of {\cite{st80}} . Although we also applied 
the correction for the motion of the spacecraft, it turned out to be 
insignificant for our analysis. Even at its maximum, the scale was less than 
1 km~s$^{-1}$ for Capella's location on the sky. Barycenter-corrected 
radial velocities $V_{bary}$ are derived as:

$$V_{bary} = V_{obs} + C_{bary} + C_{sc}$$

\noindent
where $V_{obs}$ is the measured apparent Doppler velocity, $C_{bary}$ and 
$C_{sc}$ are the barycentric correction and the correction for the motion
of the spacecraft, respectively.

\begin{figure}
\plotone{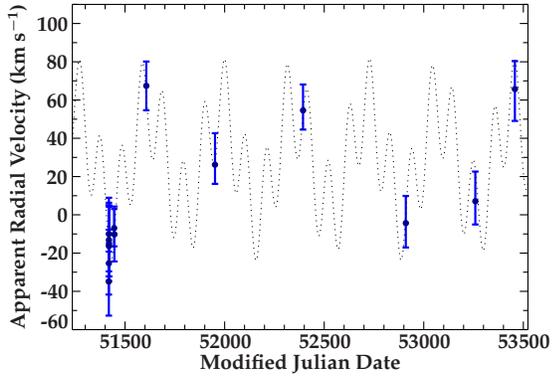}
\caption{Apparent radial velocities of Capella measured with the Chandra/HETGS. The dotted line shows the calculated apparent radial motion of Capella Aa 
viewed from Earth (including the Barycentric, orbital and systemic motion of 
Capella Aa). 3$\sigma$ error bars are shown in the plot. \label{fig1}}
\end{figure}

\begin{figure}
\plotone{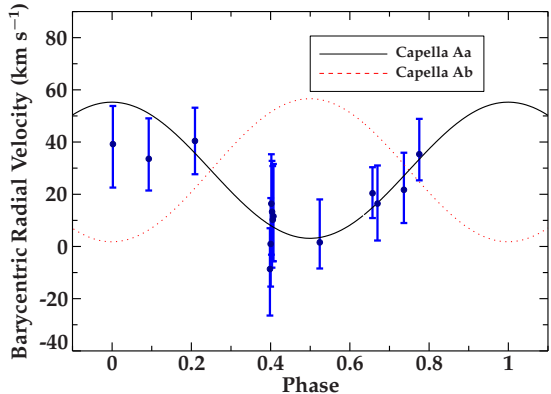}
\caption{The observed radial motion of Capella vs. orbital phase after 
barycentric correction (see the 7th column in Table~1). The measured 
radial velocity clearly follows the trend of Capella Aa (primary). 
3$\sigma$ error bars are shown in the plot.\label{fig2}}
\end{figure}

\begin{figure}
\plotone{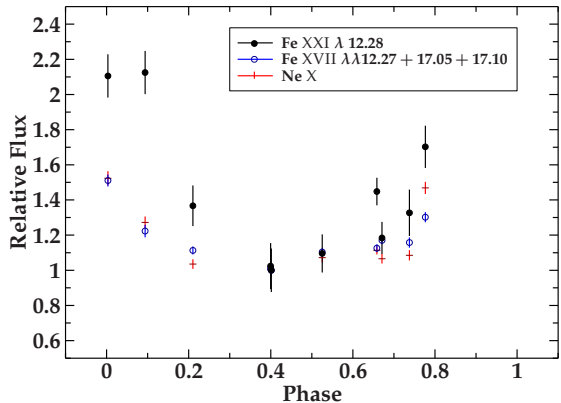}
\caption{Changes in relative integrated line flux of hot (Fe~{\sc{XXI}}; filled circle) 
and cool (three Fe~{\sc{XVII}} and Ne~{\sc{X}}; open circle and cross) emission lines. 
The measured line fluxes are scaled relative to the lowest flux observed near phase = 0.4. 
1$\sigma$ error bars are shown in the plot.\label{fig3}}
\end{figure}

\section{Detection of the Doppler Motion of Capella}

Figure~1 shows the apparent radial velocities of Capella measured 
without barycentric correction. The data points are seemingly 
randomly scattered, though close comparison with the calculated 
apparent radial motion of {\it the primary Capella Aa} (the dashed curve) 
indicates that the observed hard X-ray emission lines of Capella trace 
the primary star's motion. The zero orbital phase of the Capella system
applied in this work is defined by Hummel et al. (1994) as

$$ P_{0}(n) = JD 2447528.45 (\pm 0.02) + 104.022 \times n $$

\noindent
where $n$ is an integer. Other orbital parameters are defined as follows:
the orbital inclination $i = 137.18$\arcdeg, eccentricity $e = 0.0$, 
the masses of the primary and companion stars $M_{p} = 2.69~M_{\odot}$
and $M_{s} = 2.56~M_{\odot}$, the node of ascension $\Omega (2000.0) = 40.8$\arcdeg, 
the velocity amplitudes $K_1$ and $K_2$ for the primary and secondary stars 
are 26.05 and 27.40~km~s$^{-1}$, and the systemic velocity\footnote{the systemic 
velocity $\gamma$ and velocity amplitudes $K_1$ and $K_2$ are quoted from Barlow et al. (1993).}
$\gamma = 29.20$~km~s$^{-1}$. In Figure~2, 
the same data points were barycenter corrected and mapped into and  
plotted versus orbital phase.  The velocity vs.\ 
phase diagram makes it clearer that the X-ray emission lines detected with the 
{\it Chandra}/HETGS closely trace the motion of Capella's primary star, as 
opposed to the secondary. 

\begin{figure}
\epsscale{.80}
\plotone{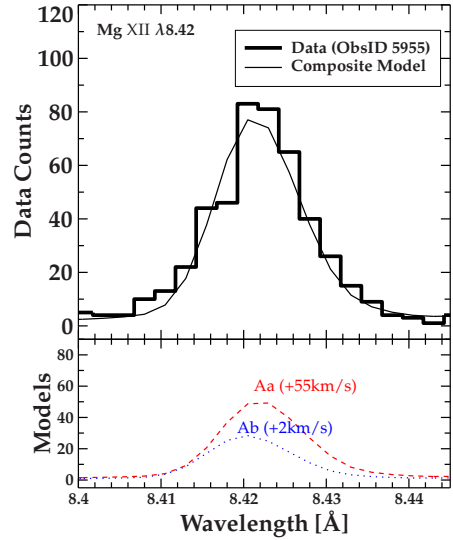}
\caption{Two-velocity-component fit to the Mg~{\sc{XII}} blend at 8.42\AA. 
At the phase = 0.002 (ObsID 5955), barycenter-corrected radial velocities for Capella~Aa 
and Ab are expected to be 55 and 2 km~s$^{-1}$, respectively. With the 
line positions fixed, the ratio of the normalization terms for Capella~Aa and
Ab is then derived to be roughly 2 : 1. \label{fig4}}
\end{figure}

\section{Comments on Science and Calibration}

In six years of monitoring Capella, the {\it Chandra}/HETGS data 
reveal that the Doppler shifts of the bright X-ray emission lines closely 
follow the primary star of the Capella system. This implies that the G8~{\sc{III}} 
primary star has been the dominant X-ray source of 10$^{7}$~K plasma, at least, 
in the last six years.

The identity of the dominant source of hot plasma in the Capella 
system has been somewhat controversial. It was long suspected that the G8 giant 
was responsible for the hard X-ray plasma, since the secondary star belongs to 
the class of X-ray deficient stars (late F -- early G giants; see Ayres et al. 1998). 
This view is strongly supported by a {\it Far Ultraviolet Spectroscopic Explorer} 
observation of Capella, which concluded that the far UV Fe~{\sc XVIII}~$\lambda$974.85 flux 
(formed at 10$^{6.8}$~K) originated solely from the G8 star's corona (Young et al. 2001). 
The evidence pointing to the slower-rotating G8 star as the dominating
source of coronal emission is particularly interesting, considering that 
the G1 star appears to dominate the UV emission formed at lower temperatures 
($T < 10^{6}$~K in the transition region and chromosphere; see Woods \& Ayres 1995 
and Linsky et al. 1995). On the contrary, the detection of strong Fe~{\sc XXI} 
$\lambda$1354 (formed at 10$^{7}$~K) far-UV emission from the G1 star 
with the {\it HST}/STIS by Johnson et al. implies that the G1 giant might 
have dominated the hotter plasma (c.f. Linsky et al. 1998).  
Variability in luminosity is another issue for concern, 
as Capella (notably the G8 component) is known to vary in the hottest UV emission 
lines (Fe~{\sc{XXI}} to {\sc{XXIV}}; see Brickhouse et al. 2000 and Ayres et al. 2003). 
In X-rays,  however, it is considered to be a steady source with 
no prior detection of flares (e.g., see Argiroffi et al. 2003). 
Is it then reasonable to infer that the two giants in Capella are 
remarkably steady X-ray sources? Light curves of Fe and Ne emission lines in Figure~3 
illustrate that this is not necessarily the case. Based on the integrated 
line fluxes measured in Fe~{\sc{XXI}} $\lambda$12.28, three Fe~{\sc{XVII}} 
$\lambda\lambda$12.27+17.05+17.10, and Ne X Ly$\alpha$ $\lambda12.13$ lines, 
it appears that the line flux from the cooler plasma (i.e., Fe~{\sc{XVII}} 
and Ne~{\sc{X}} lines) changes by 20 -- 50\% over time, while the hottest emission
line (Fe~{\sc{XXI}}) is more modulated (by a factor of two) than the cooler
counterpart. 

Furthermore, the two largest modulations of Fe~{\sc{XXI}} in 
line flux were observed in the last two observations of Capella (MJD = 53258.924
and 53457.553, or phase = 0.092 and 0.002, respectively). Examining 
Figure~2 closely, we notice that these two barycentric radial velocities are significantly 
lower than expected. It is suggestive that the deviation of the last two data points 
is probably due to an increasing level of contamination by the secondary's 
line emission flux originating from hot 10$^{7}$~K plasma. 
This scenario is supported by the observed characteristics of thermal
distribution of plasma in single giants analogous to the components of
Capella. The clump K0 giant $\beta$~Cet, analogous to Capella Aa, shows 
a thermal distribution sharply peaked at $T \sim 10^{6.8}$~K (Ayres et al. 1998); 
on the other hand, the Hertzsprung gap G0 giant 31~Com, analogous to Capella Ab, is characterized 
instead by a temperature distribution steeply increasing up to the $10^7$~K peak 
temperature and a large amount of plasma at even hotter temperatures (Scelsi et al. 2004).
Assuming that the Capella components have analogous thermal structures,  
it is then reasonable to assume that the G8 star is the stronger X-ray source, 
but the G1 star significantly contributes to the hotter emission at $T \gtrsim 10^7$~K.
Figure~4 shows that {\it indeed} bright and isolated Mg~{\sc{XII}} Ly$\alpha$ doublets can be 
fit well with two line components associated with Capella Aa and Ab.  
In the lower panel of Figure~4, the dashed line centered at barycenter-corrected radial 
Doppler velocity $+$55~km~s$^{-1}$ is the Mg~{\sc{XII}} doublets from Capella Aa (G8~{\sc{III}})
and the dotted line centered at $+$2~km~s$^{-1}$ is the same doublets from 
Capella~Ab (G1~{\sc{III}}). The composite of the two models is shown as the solid line 
in the top panel of Figure~4. In this preliminary analysis, the intrinsic line widths
for both components were set to zero. To quantify the velocity perturbation 
in terms of how much flux would be required from the secondary star to 
offset the radial velocity, the normalization terms for each line
component were allowed to vary. The best fit resulted from the line flux ratio 
(G8/G1) of 1.8 : 1 in the Mg~{\sc{XII}} feature. The weighted mean for the 
two Doppler velocities is $(55~km/s \times 1.8 + 2~km/s) / (1.8 + 1) = 36.1~km/s$, 
which is reasonably close to the measured radial velocity 
of 39.2~km~s$^{-1}$ (see Table~1). This finding indicates that the 
secondary star may have contributed as much as one third of 
the flux in the HETGS wave-band during the observations 
for ObsID 5040 and 5955. This analysis of variation in 
integrated line flux is still very preliminary and a further discussion 
is deferred to a future publication.

Lastly we find that the {\it Chandra}/HETGS  appears to be a very stable instrument 
in the long run. The improvements in the ACIS-S geometry values and the MEG grating period 
(provided in CALDB 3.0.1, version {\tt geomN0005}) further enhance its capability to 
detect a miniscule dynamic motion of an astrophysical source. The differences 
between the measured and expected Doppler velocity of Capella~Aa generally lie
under 20 km~s$^{-1}$, which we shall consider as the systematic uncertainty in 
Doppler velocity determination using the entire coverage of the  HETGS. 
As for absolute wavelength determination (i.e., per emission line), inspection
of our detailed fits indicates that the {\it r.m.s.} of the differences derived per 
emission line are 25 and 33~km~s$^{-1}$ with the HEG and MEG, respectively.

\acknowledgments
We would like to thank our referee for his insightful comments on 
this letter. We are also grateful for technical support provided by J. E. Davis and 
J. Houck at MIT. This research was supported by NASA through the SAO contract SV3-73016 
to MIT for support of the Chandra X-Ray Center and Science Instruments, 
operated by SAO for and on behalf of NASA under contract NAS8-03060.

{\it Facilities:} \facility{CXO (HETGS)}

\end{document}